\documentclass[prl,twocolumn,showpacs,amsmath,amssymb,amsfont,floatfix]{revtex4}

\usepackage{graphicx}

\begin{document}

\title{Possible observation of phase separation near a quantum phase transition
in doubly connected ultrathin superconducting cylinders of aluminum}

\author{H. Wang}

\author{M. M. Rosario}
 \altaffiliation[Present address: ]{Department of Physics and Astronomy, 
Saint Mary's College of California, Moraga, CA 94556.}

\author{N. A. Kurz}

\author{B. Y. Rock}

\author{M. Tian}

\author{P. T. Carrigan}

\author{Y. Liu}

\affiliation{Department of Physics, Pennsylvania State
University, University Park, PA 16802}

\date{\today}

\begin{abstract}

The kinetic energy of superconducting electrons in an ultrathin, doubly
connected superconducting cylinder, determined by the applied flux, increases
as the cylinder diameter decreases, leading to a destructive regime around
half-flux quanta and a superconductor to normal metal quantum phase
transition (QPT). Regular step-like features in resistance vs. temperature
curves taken at fixed flux values were observed near the QPT in ultrathin
Al cylinders. It is proposed that these features are most likely resulted
from a phase separation near the QPT in which normal regions nucleate in a
homogeneous superconducting cylinder.

\end{abstract}

\pacs{74.78.Na, 74.25.Fy}

\maketitle

%introduction

The existence of a destructive regime near half-flux quanta in ultrathin,
doubly connected superconductors \cite{degennes81} provides a rather unique
mechanism for destroying superconductivity in samples with restricted geometry.
In a doubly connected system, such as a ring, fluxoid quantization demands that
the equilibrium-state superfluid velocity, $v_s$, satisfy the relation
$v_s=\frac{2\hbar}{m^*d}(p-\frac{\Phi}{\Phi_0})$, where $\hbar$ is
Planck constant, $m^*$ is the effective Cooper pair mass, $d$
is the ring diameter, $\Phi$ is the applied flux, $\Phi_0$ (=
$h/2e$) is the flux quantum, and $p$ is an integer minimizing
$v_s$ \cite{little62}. As $d$ decreases, $v_s$ increases.
When $d$ becomes smaller than the zero-temperature
superconducting coherence length, $\xi(0)$, the kinetic energy of the
superconducting electrons near half-flux quanta will exceed
the superconducting condensation energy, leading to the
suppression of superconductivity by increasing
kinetic rather than repulsive interaction energy \cite{degennes81}.
Experimentally, this prediction was confirmed in a
cylindrical geometry because $\xi(0)$ tends to be larger in a cylinder
than in a ring when the two have similar diameters \cite{liu01}.

Cylinders possessing a destructive regime may be considered as
an one-dimensional (1D) system. Motivated by an early experiment on
possible observation of quantum phase slips \cite{giordano88}, 1D
superconductors have been studied extensively experimentally in recent
years using singly connected nanowires 
\cite{xiong97,bezryadin00,lau01,michotte04,tian05}.
Our work on doubly connected
1D superconductors has been focused primarily on destructive-regime
physics, in particular, the existence of a novel
quantum phase transition (QPT) tuned by kinetic
energy, as opposed to wire thickness \cite{bezryadin00},
and the exotic normal state in the destructive regime \cite{vafek05}.
The kinetic energy in cylindrical samples is determined by
applied flux, which can be controlled rather precisely,
allowing detailed measurements on this QPT. In
this Letter we report the observation of a possible phase separation
in a homogeneous 1D superconducting state near the destructive regime
in ultrathin Al cylinders.

Ultrathin cylinders of Al were fabricated by evaporating Al onto rotating
quartz filaments, pulled from quartz melt and attached to a thin glass slide
with a notch cut out. The Al film thickness was measured by a
quartz crystal thickness monitor during deposition. The
diameter of the cylinder was calculated from the period in resistance
oscillation, using a natural period of $h/2e$ in magnetic flux, as
done previously \cite{little62}. A scanning electron microscopy (SEM) study
was performed in several cases after all low-temperature measurements
had been carried out to measure the diameter directly and
check the structural homogeneity of the cylinder. The two methods of
determining diameter yielded similar values. Different from the previous
work, we have pursued 4-wire measurements by attaching fine Au wires
directly to the cylinder spanning the gap of the glass slide using Ag epoxy or paste,
as opposed to using films evaporated on the glass slide as 
the only measurement leads. The length between the voltage leads defined by two Ag
epoxy or paste dots is usually more than 100 $\mu$m.
Unfortunately, electrical contacts with the ultrathin cylinder do not always
survive to low temperatures, in which case two-wire measurements were carried
out. Cylinders were manually aligned to be parallel to the magnetic field.
Electric transport measurements were carried out in a dilution refrigerator
equipped with a superconducting magnet with a base temperature below 20 mK.
All leads entering the measurement enclosure were
filtered by RF filters working at room temperatures.

\begin{figure}

%\resizebox{0.45\textwidth}{!}{

\includegraphics{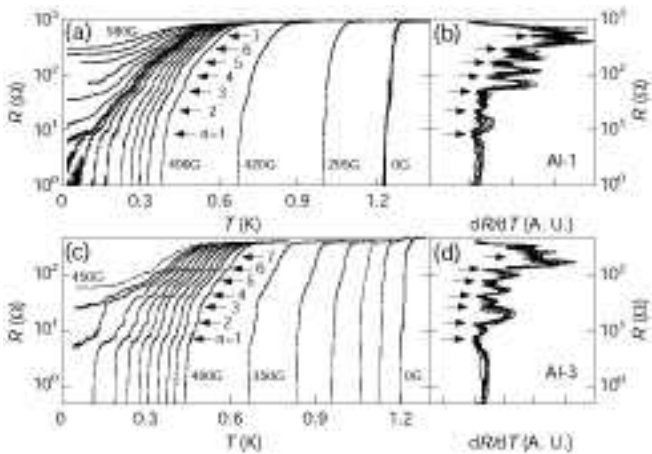}%}

\caption{\label{fig1} a) Resistance as function of temperature,
$R(T)$, at different applied flux values for Cylinder Al-1,
steps in $R(T)$ are marked by arrows. $n$
denotes the order for the step to appear as the temperature is
increased. Between 490 G and 580 G,
the field values are 500, 505, 510, 515, 520, 524, 526,
527, 528, 530, 535, 540, 550, and 560 G.
$\Phi_0$/2 corresponds to 585.5 G; b) $dR/dT$ for data shown in
1a. Steps in $R(T)$ shown in 1a correspond to minima in
$dR/dT$; c) $R(T)$ curves for Al-3. Between 400 G and 450 G, the
field values are 405, 410, 415, 420, 425, 430, 435, 440, 441.4,
and 445 G. $\Phi_0$/2 corresponds to 580 G; d)
$dR/dT$ for data shown in 1c. The bias currents were 25 nA
for Al-1 and 100 nA for Al-3.}

\end{figure}

Figures 1a and 1c show the resistance as a function of temperature,
$R(T)$, at different applied flux values from 0 to $\Phi_0$/2
for Cylinders Al-1 and Al-3, respectively. Parameters for these and
other cylinders used in this study are summarized in Table I. Cylinder
Al-1 was one of the samples used in the original experiment on the
destructive regime \cite{liu01}. As the system approaches $\Phi_0$/2,
regular step-like features, identified alternatively as minima in
$dR/dT$, are seen at fixed resistance values. At low fields, these features
become less distinct, and disappear when the field is sufficiently small.
Even though a single step is seen at zero fields in Al-1, Al-3, and
several other samples we measured, it always disappears at slightly higher
fields. Although the precise physical origin for this step is not known, we
believe that it is a sample-specific feature unrelated to those seen in
higher fields. The regular step-like features near the half-flux quantum
are clearly induced by magnetic field.

\begin{table}

\caption{\label{tab1} Parameters of cylinders reported in this work.
$\xi(0)$ is estimated from the parallel upper critical field,
$H_{c2}^\|$; the length of the cylinder, $L$, is defined by the
distance between the edges of the adjacent Ag epoxy or paste dots; the
mean-free path, $l_{el}$, is estimated from the normal-state
resistivity, $\rho_N$, using free-electron gas model. $m. c.$ denotes
measurement configurations (2- {\it vs.} 4-wire).}

\begin{ruledtabular}

\begin{tabular}{ccccccc}
      & $m. c.$ & $d$ (nm) & $t$ (nm)& $L$ (mm) & $\xi(0)$ (nm) &
$l_{el}$ (nm)\\ \hline
     Al-1 & 2 & 150 & 30 & 0.43 & 161 & 13\\
     Al-3 & 2 & 151 & 33 & 0.29 & 190 & 16\\
     Al-4 & 2 & 157-169 & 31.5 & 0.48 & $\approx $ 97 & 6.1\\
     Al-5 & 4 & 212 & 30 & 0.11 & 146 & 7.9\\
     Al-6 & 4 & 263 & 30 & 0.11 & 150 & 13\\
\end{tabular}

\end{ruledtabular}

\end{table}

Even though both samples shown in Fig. 1 were measured in a 2-wire
configuration, the step-like features are not due to 2-wire measurements
because of the following reasons. First, these
features were seen in both 2- and 4-wire samples (see below). Furthermore,
in a control experiment, 2- and 4-wire measurements were carried out on
the same cylinder with multiple leads ($d \approx 0.2 \mu$m). Essentially
identical step-like features in $R(T)$ were found in both cases. Sample
inhomogeneities featuring variation in local $T_c$ can also be excluded
from being the cause of the regular step-like features. To begin with,
SEM studies have shown that our cylinders are quite
uniform. For those on which sharp SEM images were obtained,
grains of Al with rather uniform size were seen. In addition, if the sample
were inhomogeneous with separate regions of different $T_c$ values, the
superconducting transition in zero fields would not be as sharp as what was
observed experimentally. The regular step-like features would also have
persisted down to low and zero fields, inconsistent with experimental result.

Can the step-like features be due to phase slip centers (PSCs) formed at sections
of the sample with a small local critical current ($I_c$) \cite{SBT74}?
We believe that they are not based on the following observations. First,
the detailed evolution of the step-like features as a function of
magnetic field suggest that the step-like
features were not caused by PSCs; Second, we measured a
cylinder (Al-3) with different bias currents. It appears that the step-like
features are present even at 10 nA. More importantly, at high bias currents,
some step-like features presented at lower currents actually disappeared,
again inconsistent with the PSC picture; Finally, we measured
$R(T)$ at fixed magnetic field with a battery rather than a digital current
source that may have introduced electrical noises to the sample despite the
damping of the RF filters. Since electrical noises are known to generate PSCs,
stronger step-like features are expected for digital current source 
measurements
if these features were due to PSCs. Experimentally, the opposite is true.

\begin{figure}

%\resizebox{0.45\textwidth}{!}{

\includegraphics{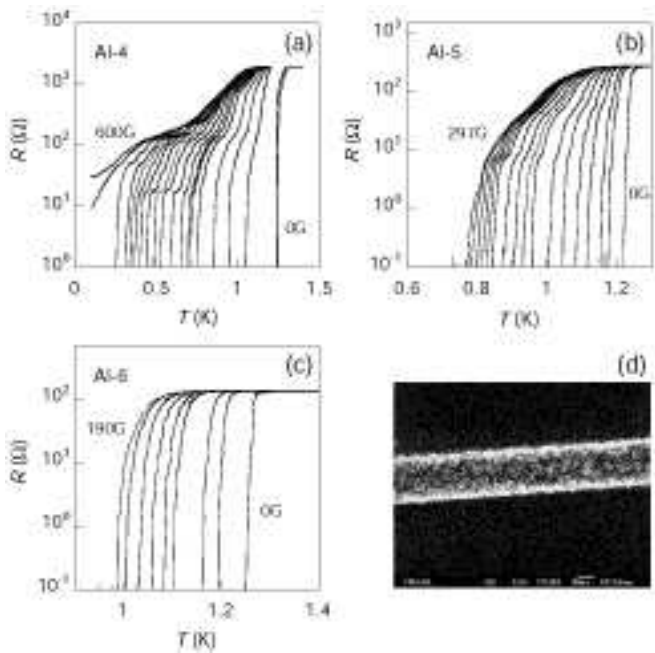}%}

\caption{\label{fig2} a-c) $R(T)$ curves at
various flux values for cylinders with a larger diameter (Table I).
The magnetic field corresponding to $\Phi_0$/2 is 460-533 G
for Al-4 which is slightly non-uniform in diameter, 295 G for Al-5, and
190 G for Al-6. The measurement current was 100 nA for all three cylinders;
d) An SEM picture for Al-6. The Al grains are seen to be quite uniform
with an average size of 69$\pm$16 nm. The
diameter measured in the SEM image is 268$\pm$10 nm, as compared to
263$\pm$7 nm obtained from the $R(H)$ measurements.}

\end{figure}

Figure 2 shows $R(T)$ traces at fixed flux values up to half-flux quantum
for cylinders with a diameter slightly larger than that required for
destructive regime (Table I). Multiple, but relatively irregular step-like
features were observed in Al-4, which is only slightly larger than Al-1 and
Al-3. The irregularity in the step-like features appears to be related to the
slight variation in diameter (Table I) as revealed by the $R(H)$ measurements.
The overall trend of the step-like features are very similar to those found
In Al-1 and Al-3. In Al-5, which is larger in diameter but uniform, only a
couple of step-like features were found. For Al-6, with the largest diameter
(Fig. 2d), no step-like features were observed. This
systematic behavior was observed for $all$ cylinders (9 in total) in
which the presence of such step-like features was examined closely,
suggesting that the step-like features can be induced by magnetic field
so long as the diameter of the cylinder is sufficiently close to
that required for possessing a destructive regime.

\begin{figure}

%\resizebox{0.45\textwidth}{!}{

\includegraphics{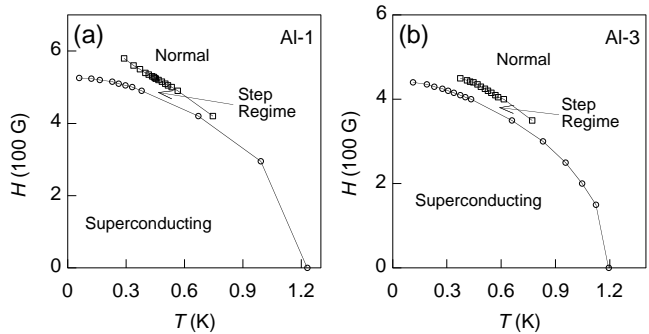}%}

\caption{\label{fig3} Empirical phase diagram for
Al-1 and Al-3. The upper curve marks the highest $T$ at which
a step was identified at fixed flux while the lower curve shows
the onset of finite resistance.}

\end{figure}

For the two samples shown in Figs. 1a and b, the empirical $\Phi-T$
phase diagrams are constructed (Fig. 3a and b). The phase
diagrams suggest that for cylinders with sufficiently small diameters to host
a destructive regime, the step-like features were found near the QPT between
the superconducting and normal ground states at $T=0$. If we consider the
diameter of the cylinder as the third axis in the parameter space, the above
results seem to suggest that the step-like features emerge in a quantum
critical regime. It is interesting to note that the appearance of the
step-like features shown in Figs. 1 and 2 is accompanied by a broadening
of $R(T)$, a feature very similar to that observed near a
superconductor-insulator transition (SIT) in 2D \cite{liu93} homogeneous
systems. Even though whether the superconductor-normal metal transition
at onset of the destructive regime is a continuous QPT featuring a critical
regime is yet to be established, it is likely that whatever drives this QPT
is also responsible for the emergence of the regular step-like features.

In 2D SIT, the QPT is widely believed to be driven by phase
fluctuation due to the suppression of fluctuation in the number of Cooper
pairs, $N$. Because of the relation, $\Delta N\Delta \phi>1$, where $\phi$ is
the phase of the superconducting order parameter, the suppression in
$\Delta N$ enhances $\Delta \phi$. However, $\Delta N$ is not suppressed near
the destructive regime. Therefore it seems that even though the phase
fluctuation may be present because of the reduced dimension, it can not
dominate the QPT. A different path must be explored in order
to understand this QPT and the accompanied step-like features. In this regard,
the remarkable regularity of the latter as revealed in Fig. 1 may provide
useful hints. To quantify this regularity, in Fig. 4a, we plot $R_{step}$,
at which the
step-like features were found, as a function of $n$, which denote the order
for the emergence of the step-like feature as $T$ increases (Fig. 1). We see
clearly that $\log_{10}R_{step} \propto n$, namely, $R_{step}$ grows
exponentially.

\begin{figure}

%\resizebox{0.45\textwidth}{!}{

\includegraphics{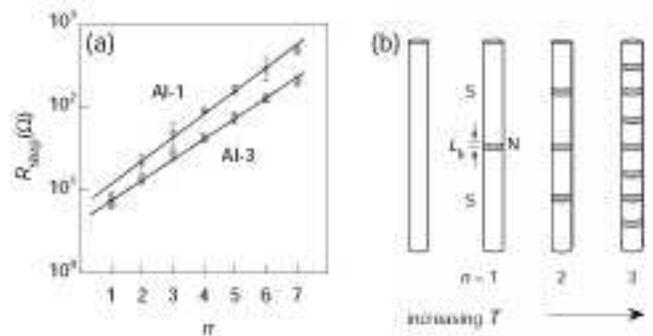}%}

\caption{\label{fig4} a) Step resistance as function of the index
$n$ (Fig. 1) for Al-1 and Al-3. The lines indicate exponential dependence;
b) Proposed bifurcation process for the normal band
formation in ultrathin cylinders.}

\end{figure}

To explain such exponential growth in $R_{step}$, and the emergence of the
step-like features, we propose the following phenomenological model.
As the system approaches the destructive regime, normal regions will nucleate
in a homogeneous superconducting state. Each normal region will encircle the
entire cylinder, forming a ring, or a band, referred here as a normal band.
The first normal band will form near the center of a uniformly
superconducting cylinder as $T$ is raised. With increasing $T$, each of the
two superconducting sections will break into two sections, followed by breaking
the 4 superconducting sections, and so on (Fig. 4b). Such a bifurcation
process can lead to an exponential growth in the number of normal bands, $N$,
given by $N = 2^n-1$. If all normal bands have the same length,
we have $R_{step} = N R_1$, where $R_1$ is the resistance associated with a
single normal band, leading to evenly spaced steps on a logarithmic scale as
seen experimentally.

Further analysis shows good self consistency in this normal-band model.
The slope in Fig. 4a is only slightly smaller than expected $\log_{10}2$.
The value of $R_1$ is also consistent
with that expected for a single normal band. The length of a normal band,
$L_b$, should be $2\xi(0)$ based on energetic considerations (see below).
However, $R_1$ should correspond to the resistance of a length twice of
$\Lambda_Q$, the charge imbalance length \cite{clarke72}. Typically $\xi(0)
<< \Lambda_Q$. Therefore, $R_1=2\Lambda_Q\cdot\rho_N/A$, where $A$ is the
cross-section area, similar to that for PSCs \cite{SBT74}. Based on the
experimental value of $R_1$, $\Lambda_Q\approx 2\mu m$ for Al-1 and 3, a
very reasonable number for Al \cite{strunk98,kadin78}.
In principle, $\Lambda_Q$ is a function of the applied field and temperature
\cite{kadin78,michotte04}. Therefore $R_1$ at different $\Phi$ values should
be different. However, our calculations show that $\Lambda_Q$ varies within
10\% for all curves with step-like features in Fig. 1. Such a variation
in $R_1$ is invisible in a logarithmic plot.

Even though this picture of normal-band bifurcation seems to provide a
consistent account of our data, it is surprising that a regular spatial
variation of the order parameter should be allowed as this would in general
cost energy. On the other hand, as the destructive regime is approached, the
free energy of the normal state is only slightly higher than that of the
superconducting state because of the large $v_s$. To minimize the free energy
cost, a normal band should only be long enough to support two
superconducting-normal (S-N) interfaces. As a result, $L_b=2\xi(0)$.
Once a normal band is
formed, two S-N interfaces should bring about an interface energy. The
applied field in this case is perpendicular rather than parallel to the
interface, different from the typical situation considered in bulk
Type I or Type II
superconductors. The energy associated with such an interface has not
been calculated. Two adjacent superconducting sections may also be coupled by
Josephson coupling, likely to lead to a gain (lowering) of free energy. All
these factors have to be considered to provide an energetic underpinning for
the normal-band formation.

It is interesting to note that steps in $R(T)$ were reported long ago in
Al cylinders with a diameter larger than or equal to 1.4 $\mu$m in a narrow
temperature range \cite{meyers71}. These cylinders were too large to
possess a destructive regime. Most importantly, all steps seen at finite
fields were also found at zero fields. Therefore the physical origin of those
steps, not identified in the original work, cannot be the same as what we have
observed in ultrathin superconducting cylinders. It was proposed previously that a 
heterogeneous
mixed state featuring isolated superconducting spots would be formed at the
high-temperature part of the superconducting transition due to impurity
and strain effects \cite{parks64}. It was further predicted that
the presence of these superconducting spots will lead to steps in $R(T)$.
No steps were actually observed in their $R(T)$ curves
taken on cylinders with a diameter larger than or equal to 1.2 $\mu$m \cite{parks64},
which were again too large to exhibit a destructive regime.

In summary, we have observed step-like features near a QPT at the onset of
the destructive regime in ultrathin, doubly connected superconducting cylinders.
A tentative model based on phase separation is proposed to explain 
the emergence of these step-like features. More theoretical input and further 
experimental studies will be needed to fully understand the physical origin of 
these step-like features. For example, reducing the wall thickness of the cylinder 
will increase the amount of disorder and decrease the orbital effects. Varying
the magnetic field angle with respect to the axis of the cylinder, which were found
to give rise to Abrikosov vortices in a large cylinder, may shed light on the microscopic 
origin of the proposed phase separation near the QPT in our ultrathin cylinders.

The author would like to thank useful discussions with Professors M.Beasley,
H. Fan, M. Ma, M. Sigrist, M. Tinkham and Drs. Oskar Vafek and Z. 
Long. The work
is supported by NSF through grants DMR-0202534 and DMR-0080019.

\end{document}